\documentclass{PoS}

\usepackage[utf8]{inputenc}

\usepackage[english]{babel}
\usepackage{amsthm, amssymb, amsmath, latexsym}
\usepackage{graphicx, graphics, subfigure}
\usepackage{multirow}
\usepackage{slashed}

\usepackage{xcolor}
\usepackage{float}
\usepackage{cite}

\newlength{\bredde}
\def\slash#1{\settowidth{\bredde}{$#1$}\ifmmode\,\raisebox{.15ex}{/}
\hspace*{-\bredde} #1\else$\,\raisebox{.15ex}{/}\hspace*{-\bredde} #1$\fi}

\def\amp{{\mathcal{M}}}

\def\beq{\begin{equation}}
\def\bea{\begin{eqnarray}}
\def\eeq{\end{equation}}
\def\eea{\end{eqnarray}}

\def\caliD{{\mathcal{D}}}

\title{Investigation of the phase structure of a chirally-\\invariant 
	Higgs-Yukawa model}
\ShortTitle{Phase structure of a chirally-invariant 
	Higgs-Yukawa model}

\author{John~Bulava$^{a}$, Philipp~Gerhold$^{b,c}$, George~W.-S.~Hou$^{d}$,
Karl~Jansen$^{c}$, Bastian~Knippschild$^{d}$,
C.-J. David~Lin$^{e,h}$, Kei-Ichi~Nagai$^{f}$, \speaker{Attila~Nagy}$^{\,\,b,c}$, Kenji~Ogawa$^{g}$\\\\
	$^{a}$ CERN, Physics Department, 1211 Geneva 23, Switzerland\\
	$^{b}$ Institut f\"ur Physik, Humboldt-Universit\"at zu Berlin, D-12489 Berlin, Germany\\
	$^{c}$ NIC, DESY, Platanenallee 6, Zeuthen D-15738, Germany\\
	$^{d}$ Department of Physics, National Taiwan University, Roosevelt Road, Taipei 10617, Taiwan\\
	$^{e}$ Institute of Physics, National Chiao-Tung University, Hsinchu 300, Taiwan\\
	$^{f}$ Kobayashi-Maskawa Institute, Nagoya University, Nagoya, Aichi 464-8602, Japan\\
	$^{g}$ Department of Physics, Chung-Yuan Christian University, Chung-Li 32023, Taiwan\\
	$^{h}$ Division of Physics, National Centre for Theoretical Sciences, Hsinchu 300, Taiwan\\\\
        	E-mail: \email{nagy@physik.hu-berlin.de}\\
	E-mail: \email{b.knippschild@gmx.de}\\\\}

\abstract{We present new data on our ongoing project on the investigation of the phase structure of the Higgs-Yukawa model at large bare Yukawa couplings. The data presented last year~\cite{Bulava:2011jp} are extended in terms of statistics, the number of bare Yukawa couplings at existing, and new larger volumes. In addition, this study is extended by a finite temperature project at the physical top quark mass $m_t =175\text{ GeV}$ and a hypothetical fourth generation top quark with a mass of $m_{t'}=700\text{ GeV}$.}

\FullConference{The 30 International Symposium on Lattice Field Theory - Lattice 2012,\\
		June 24-29, 2012\\
		Cairns, Australia}

\begin{document}

%%%%%%%%%%%%%%%%%%%%%%%%%%%%%%%%%
%
%						Introduction
%
%%%%%%%%%%%%%%%%%%%%%%%%%%%%%%%%%
\section{Introduction}
Standard Model Electroweak Baryogenesis has difficulties explaining the matter-antimatter asymmetry present in the universe. A first-order phase transition is required to fulfil one of  the Sakharov conditions, but the end point of the first-order phase transition line occurs at an unphysical low Higgs boson mass~\cite{Kajantie:1996mn, Fodor:1994sj}. The transition is first order in the Standard Model and second order in the pure $\phi^4$ theory, but the situation is less clear when fermions with strong Yukawa coupling are added, in particular when a heavy fourth generation of quarks are considered~\cite{Hung:2010xh, Kikukawa:2009mu}.

\smallskip
It is known that the Higgs-Yukawa model approaches the non-linear $\sigma$-model at asymptotically large Yukawa couplings~\cite{Hasenfratz:1988vc}, but it is not known where this behaviour sets in. We investigate the bulk phase transition of the Higgs-Yukawa model non-perturbatively at large bare Yukawa couplings with lattice techniques at finite and zero temperature. In contrast to previous calculations~\cite{Bock:1991bu}, we consider a chirally invariant fermion action~\cite{Neuberger:1998wv} with a complex scalar doublet $\phi$ and a fermion doublet $\psi = \left(t, b \right)^T$. We wish to investigate the possibility of an intermediate region in the Yukawa coupling where the ultraviolet behaviour may not be trivial anymore and a heavy fourth generation of quarks is possible.

%%%%%%%%%%%%%%%%%%%%%%%%%%%%%%%%%
%
%					Simulation details
%
%%%%%%%%%%%%%%%%%%%%%%%%%%%%%%%%%
\section{Simulation details}

The discretised action of a four-component scalar field theory with quartic self interaction is
\beq
\label{eq:scalar_action_ising_form}
  S_{\phi} = - 2\kappa 
      \sum_{x,\mu} \phi^{\alpha}_{x} \phi^{\alpha}_{x + \hat{\mu}}
       + \sum_{x} \left [ \phi^{\alpha}_{x}\phi^{\alpha}_{x} + 
        \hat{\lambda}(\phi^{\alpha}_{x}\phi^{\alpha}_{x}-1)^{2}\right ] .
\eeq
where $\alpha$ labels the four components of the scalar fields, $\kappa$ is the so-called hopping parameter which is related to the Higgs bosons mass, $\hat\lambda$ is related to the bare quartic self-coupling, and the lattice spacing $a$ is set to $1$. The sum is performed over all space time points, $x$, and four dimensions, $\mu$. For the lattice fermion action which includes the Yukawa interaction of the Higgs $vev$ and the fermions, we use the overlap operator $\caliD^{({\mathrm{ov}})}$ 
\beq
\label{eq:fermion_action}
 S_{F} = \bar{\Psi} \amp \Psi , \mbox{ }{\mathrm{where}}\mbox{ }\mbox{ }
  \amp = \caliD^{({\mathrm{ov}})} 
    + P_{+} \mbox{ }\Phi^{\dagger}\mbox{ }\mbox{ }{\mathrm{diag}}(y_{t},y_{b})\mbox{ }\hat{P}_{+}
    + P_{-} \mbox{ }\mbox{ }{\mathrm{diag}}(y_{t},y_{b})\mbox{ }\Phi\mbox{ }\hat{P}_{-} ,
\eeq
with
\beq
\label{eq:Phi_and_Psi}
\Phi = \begin{pmatrix} \phi^{0}-i\phi^{3} & -\phi^{2} - i \phi^{1} \\ \phi^{2}-i\phi^{1} & \phi^{0} + i \phi^{3}  \end{pmatrix} \, \text{, and}\,\,\,\,\, P_{\pm} = \frac{1 \pm \gamma_{5}}{2},\mbox{ }\mbox{ }
 \hat{P}_{\pm} = \frac{1 \pm \hat{\gamma}_{5}}{2}, \mbox{ }\mbox{ }
 \hat{\gamma}_{5} = \gamma_{5} \left ( 1 - \caliD^{({\mathrm{ov}})} \right ).
\eeq
The Yukawa couplings are set to $y_{t} = y_{b} = y$, to ensure that the fermion determinants are positive definite. The scalar field configurations are generated with the polynomial Hybrid Monte Carlo algorithm~\cite{Frezzotti:1997ym}, treating the weight factor as an observable \cite{Gerhold:2010wy}. Our measurements are performed on $\sim 2000 - 5000$ thermalised trajectories. 

%%%%%%%%%%%%%%%%%%%%%%%%%%%%%%%%%
%
%                        Basic observables and analysis
%
%%%%%%%%%%%%%%%%%%%%%%%%%%%%%%%%%
\section{Basic observables and analysis}
The phase structure of a theory can be investigated by measuring an order parameter which is the scalar $vev$, $v$,  in our case. It is zero in one phase, the symmetric phase, and non-zero in another phase, the broken phase, respectively. All investigated phase transitions of our model are found to be of second order where the change in $v$ is smooth. The universality classes of second order phase transitions are defined via the anomalous dimension of the operators which are allowed by the symmetries of the theory. Anomalous dimensions will be called critical exponents to draw analogy with statistical mechanics.

\smallskip

In the lattice Higgs-Yukawa model $v$ would vanish in both phases without the use of external sources which renders the computation rather demanding. However, it can be replaced by an alternative method described in Refs.~\cite{Hasenfratz:1989ux,Gockeler:1991ty} where the scalar fields are rotated 
\begin{equation}\label{eq:bare_vev_definition}
 \left<\phi_{\text{rot}}\right> = \left( \begin{array}{c} 0 \\ v \end{array} \right), \quad v=\sqrt{2\kappa} \langle m \rangle,
\end{equation}
and projected on the direction of magnetisation
\beq
\label{eq:mag}
m = \frac{1}{V_{4}} \left ( \sum_{\alpha,x}
  |\phi_{x}^{\alpha}|^{2} \right )^{1/2} \mbox{ }\mbox{ } (V_{4}
\mbox{ }{\mathrm{is}}\mbox{ }{\mathrm{the}}\mbox{ }4{-}{\mathrm{dimensional}}\mbox{ }{\mathrm{volume}}).
\eeq
The $vev$ computed in this way coincides with the scalar $vev$ in the infinite-volume 
limit. In finite volume, it can be shown that this method is equivalent to the method with external sources~\cite{Gockeler:1991ty}. 

\smallskip

It is challenging to extract the critical exponents. As mentioned before, second order phase transition are  smooth transitions from the broken to the symmetric phase and become cross-overs in a finite volume. Hence, finite size scaling techniques can be used to determine the critical exponents. We use two different methods here.

\smallskip

The first method is based on the susceptibility
\begin{equation}
\label{eq:susc_def}
	\chi_{m} = V_{4} \left( \left< m^2\right> - \left< m\right>^2,
        \right) ,
\end{equation}
which is the connected two-point function and it is proportional to the correlation length $\xi\propto \sqrt{\chi_m}$ which diverges in the infinite-volume limit. Close to the critical point, its finite size scaling behaviour is given by
\begin{equation}
\label{eq:susc_scaling}
	\chi_{m}\left(t, L\right)\cdot L_s^{-\gamma/\nu} =  g\left(t
          L_s^{1/\nu}\right)\text{,\, with \,} t=\left[T/\left(T_c^{(L=\infty)} - C\cdot L_s^{-b}\right) -1\right]
\end{equation}
where, $L_s$ is the spatial lattice extent, $T$ stands either for the Yukawa coupling or the hopping parameter, respectively, $T_c^{(L=\infty)} $ is its critical value in infinite volume, $g$ is an unknown universal scaling function, $\gamma$ and $\nu$ are related to the anomalous dimension of the scalar field and the mass operator, respectively, and $C$ and $b$ are phenomenological parameters.  The critical exponents can be extracted from a fit to the susceptibility via the partly-empirical fit function defined in Ref.~\cite{Jansen:1989gd}
\begin{align}
	\chi_{m} = A\left(L_s^{-2/\nu} + B_\pm \cdot t^2\right)^{-\gamma/2},
	\label{FitFunction}
\end{align}
where $A$ and $B_\pm$ are phenomenological parameters. The parameter $B_{-}$ is used when $t\le0$ and $B_{+}$ when $t>0$.

\smallskip

An alternative procedure to extract the critical exponent, $\nu$, is via Binder's cumulant \cite{Binder:1981sa}
\begin{equation}
\label{eq:binders_cumulant_def}
	Q_{L} = 1 - \frac{\left< m^4\right>}{3\left< m^2\right>^{2}} ,
\end{equation}
which is the connected four-point function, normalised by the square of the two-point function.\newpage\noindent Binder's cumulant is related to the renormalised scalar quartic coupling in the infinite-volume limit by the proportionality factor $V_4/\xi^4$ \cite{Freedman:1982zu}. The great advantage of this quantity is the milder scaling violation due to higher-dimensional operators~\cite{Privman:1983dj, Binder:1985FSTH}. To extract the critical exponent, $\nu$, from Binder's cumulant we use a method described in Ref.~\cite{0305-4470-34-33-302}. This method relies on the fact that the scaling behaviour close to the critical point of Binder's cumulant is given by
\begin{equation}
\label{eq:Binder_cumulant_scaling}
	Q_L  = g_{Q_L}\left( \hat tL^{1/\nu}\right)\text{, with }\hat t = (T/T_c-1),
\end{equation}
where $ g_{Q_L}$ is an unknown universal scaling function, which can be mimicked using part of the data.

\smallskip

The analysis for the zero temperature phase transition presented in the following assumes that the Higgs-Yukawa model is not trivial at large bare Yukawa couplings. It was shown in Ref.~\cite{Brezin:1981gm,Brezin:1985xx,Bernreuther:1987hv,Kenna:1992np,Kenna:2004cm} that logarithmical corrections may need to be included in the finite size scaling in a trivial theory. This will be investigated in the future.

%%%%%%%%%%%%%%%%%%%%%%%%%%%%%%%%%
%
%					Large Yukawa coupling
%
%%%%%%%%%%%%%%%%%%%%%%%%%%%%%%%%%
\section{Large Yukawa coupling}
%%%%%%%%%%%%%%%%%%%%%%%%%%%%%%%%%
The phase structure of the Higgs-Yukawa model is investigated at large bare Yukawa couplings. This is interesting because large renormalised Yukawa couplings would lead naturally to a heavy fourth generation even with the presence of a light Higgs Boson. In addition, it is not clear if the model is trivial in this region and hence perturbation theory is applicable. This would resolve the hierarchy problem as perturbative artefacts. To reveal the universality class we compute the critical exponents with the methods discussed previously. 

\smallskip

\begin{figure}[H]
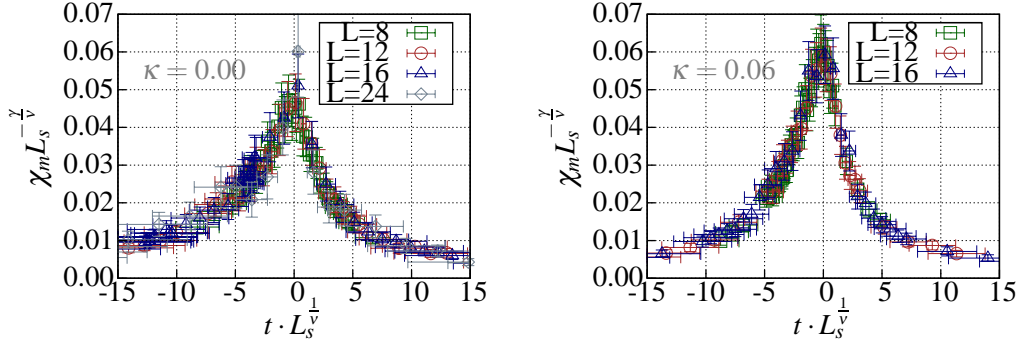

	\begin{center}
	$	
	\begin{array}{ccc}
		\hspace*{-0.5cm} \input{./sus_rescaled_kap000.tex} &
		\hspace*{-0.0cm} \input{./sus_rescaled_kap006.tex}
	\end{array}	
	$
	\end{center}
\vspace{-0.9cm}
\caption{Test of finite size scaling of eq.~\ref{eq:susc_scaling} of susceptibility.}
\label{zeroT_sus_rescaled}
\end{figure}

Our simulations have been performed at two $\kappa$ values, $\kappa = 0.00$ and $0.06$, with the bare Yukawa coupling $y$ in the range between $14$ and $25$. The bare scalar quartic coupling $\hat{\lambda}$ is fixed to infinity, which results in the largest possible Higgs mass~\cite{Gerhold:2010wv,Gerhold:2009ub,Gerhold:2010bh}.  
\vspace{-0.3cm}
\begin{figure}[H]
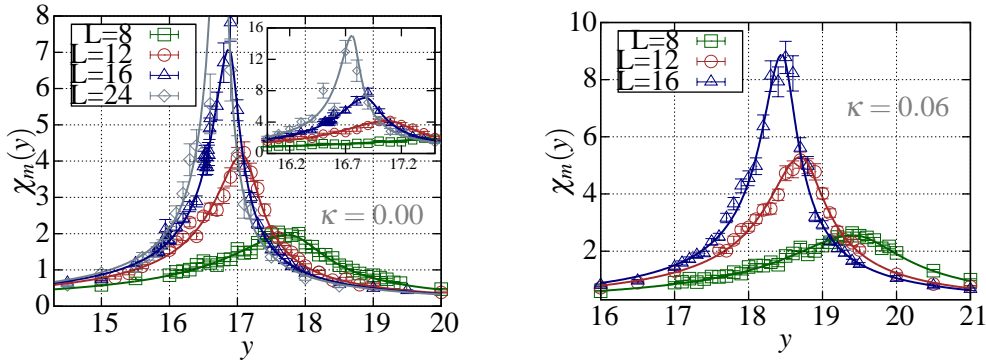

	\begin{center}
	$
	\begin{array}{ccc}
		\hspace*{-0.5cm} \input{./susceptibility_kap000.tex} &
		\hspace*{-0.0cm} \input{./susceptibility_kap006.tex}
	\end{array}
	$
	\end{center}
\vspace{-0.9cm}
\caption{Fits according to eq.~\ref{FitFunction} to susceptibility for several volumes at $\kappa=0.00$ and $\kappa=0.06$.}
\label{zeroT_susceptibility}
\end{figure} 
%%%%%%%%%%%%%%%%%%%%%%%%%%%%%%%%%
The fits to susceptibility with eq.~\ref{FitFunction} are shown in Fig.~\ref{zeroT_susceptibility} for both $\kappa$ values. A clear dependence of the peak position on the volume can be observed which is accommodated with the phenomenological parameters $C$ and $b$ in eq.~\ref{FitFunction}. The finite size scaling of susceptibility according to eq.~\ref{eq:susc_scaling} is shown in Fig.~\ref{zeroT_sus_rescaled} where the parameters are set from the fits. It is found to be very good.

\smallskip

Rescaled Binder's cumulant is shown in Fig~\ref{zeroT_susceptibility} for both $\kappa$-values after rescaling according to eq.~\ref{eq:Binder_cumulant_scaling} with the method described in Ref.~\cite{0305-4470-34-33-302}. From this method the critical exponent, $\nu$, and the critical Yukawa coupling in infinite volume can be extracted with high precision. However, the scaling is only good close to the critical point and further tests are needed to stress the interval in which this method is applicable.
\vspace{-0.5cm}
%%%%%%%%%%%%%%%%%%%%%%%%%%%%%%%%%
%%%%%%%%%%%%%%%%%%%%%%%%%%%%%%%%%
\vspace{-1.0cm}
%%%%%%%%%%%%%%%%%%%%%%%%%%%%%%%%%
\begin{figure}[H]
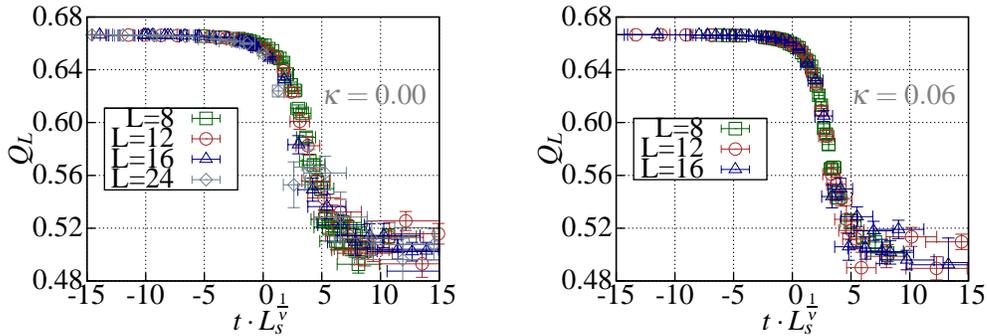

	\begin{center}
	$
	\begin{array}{ccc}
		\hspace*{-0.5cm} \input{./binder_rescaled_kap000.tex} &
		\hspace*{-0.0cm} \input{./binder_rescaled_kap006.tex}
	\end{array}
	$
	\end{center}
\vspace{-0.9cm}
\caption{Rescaled Binder's cumulant after using the  curve collapse method from Ref.~\protect\cite{0305-4470-34-33-302}.}
\label{zeroT_binder}
\end{figure} 
%%%%%%%%%%%%%%%%%%%%%%%%%%%%%%%%%
\vspace{-0.2cm}
The critical Yukawa couplings in infinite volume and critical exponents are summarised in Table~\ref{summarytable_zeroT}. All quoted errors are purely statistical but a preliminary analysis of systematic errors from the dependence of the fit interval and the interval where the curve collapse methods is used has been performed. The systematic errors of the fit are of the order of the statistical ones but the systematic errors of the curve collapse method can be five times larger than the corresponding statistical error. Within the combination of statistical and systematic errors the fit and the curve collapse method give consistent results. The critical exponent, $\nu$, is at least two standard deviations away from its trivial value of $0.5$. This deviation must be investigated further to reveal if this is a purely statistical effect, due to the neglecting of the mentioned logarithmic correction
or, more excitingly, if we found a different ultraviolet behaviour. 

%%%%%%%%%%%%%%%%%%%%%%%%%%%%%%%%%
\begin{table}[H]\centering
 \begin{tabular}{|c||c|c|c|c|c|c|}\hline
	&  \multicolumn{2}{ |c| }{ $y_c(L=\infty)$ }  & \multicolumn{2}{ |c| }{ $\nu$ }   &\multicolumn{2}{ |c| }{$\gamma$} \\ \hline
	&  $\kappa=0.00$ &  $\kappa=0.06$ &  $\kappa=0.00$ &  $\kappa=0.06$ &  $\kappa=0.00$ &  $\kappa=0.06$ \\ \hline\hline
	fit to $\chi_m$    & 16.676(15)    &  18.119(67)     &  0.541(22)     &  0.576(28)     & 0.996(15)      & 1.038(30) \\ \hline   
	method from \cite{0305-4470-34-33-302}  &  16.667(27)   &  18.147(24)   &  0.525(6)     &  0.550(1)     &  \multicolumn{2}{ |c }{ }  \\ \cline{1-5}
\end{tabular}
\caption{Preliminary results for the critical Yukawa couplings and exponents.}
\label{summarytable_zeroT}
\end{table}
%%%%%%%%%%%%%%%%%%%%%%%%%%%%%%%%%
\vspace{-0.65cm}

%%%%%%%%%%%%%%%%%%%%%%%%%%%%%%%%%
%
%				         Finite temperatur
%
%%%%%%%%%%%%%%%%%%%%%%%%%%%%%%%%%
\section{Finite temperature}
\vspace{-0.2cm}
Here we want to present first results of our finite temperature studies for a physical top quark. The simulations for a very heavy fourth generation quark at $m_f\approx700\text{ GeV}$ are in production. In particular, we are interested in the critical temperature where the system goes from a symmetric phase into a broken phase, as well as the order of the phase transition. The temperature of the system is given by
\begin{equation}
	T = \frac{1}{aL_t} = \frac{\Lambda}{L_t},
\end{equation}
where $L_t$ is the number of lattice points in temporal direction. The lattice spacing $a$ and hence the cutoff $\Lambda$ which is inversely proportional to the lattice spacing will be set in a later zero temperature run via the renormalised $vev$.

\smallskip

The finite temperature runs are performed at two fixed bare Yukawa coupling values to fix fermion masses at $m_f\approx175 \text{ GeV}$ and $m_f\approx700 \text{ GeV}$ while the scan is performed in $\kappa$ which is equivalent to varying the temperature. To investigate discretisation effects, simulations for two values of the temporal extend $(L_t=4,6)$ are performed, and to study volume effects, three different spatial volumes $(L=16, 20, 24)$ are investigated.

\smallskip

\begin{figure}[H]
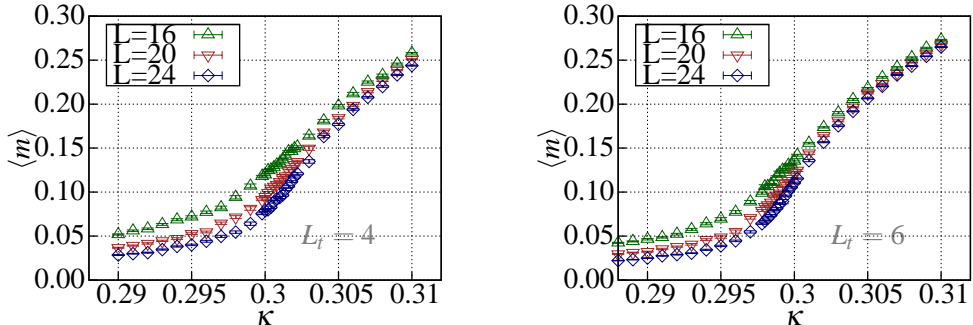

	\begin{center}
	$	
	\begin{array}{ccc}
		\hspace*{-0.5cm} \input{./finiteT_mt175_T4_mag.tex} &
		\hspace*{-0.0cm} \input{./finiteT_mt175_T6_mag.tex}
	\end{array}	
	$
	\end{center}
	\vspace{-1.2cm}
	\caption{Magnetisation at $m_f \sim 175\text{ GeV}$ for three different volumes and two temporal extends.}
	\label{fig:finiteT}
\end{figure}

The magnetisation at fermion mass of $m_f=175\text{ GeV}$ is shown in Fig.~\ref{fig:finiteT} for two different temporal extends and three volumes. A clear trend of the magnetisation from the symmetric phase with very small magnetisation to the broken phase with large magnetisation can be observed. This trend is very smooth without any jumps and hence it is a clear indication for a second order phase transition. However, further analysis of susceptibility will show if the critical exponents are consisted with the ones from mean field theory. This will tell us if this phase transition differs in any kind from the phase transition observed in pure $O(4)$-model.
\vspace{-0.6cm}
%%%%%%%%%%%%%%%%%%%%%%%%%%%%%%%%%
\vspace{-0.6cm}

%%%%%%%%%%%%%%%%%%%%%%%%%%%%%%%%%
%
%					Summary and outlook
%
%%%%%%%%%%%%%%%%%%%%%%%%%%%%%%%%%
\section{Summary and outlook}
We have found a second order bulk phase transition at large values of bare Yukawa coupling. It is not clear if the critical exponents are consistent with the trivial ones from the $O(4)$-model or if we found different ultraviolet behaviour in this regime. Possible logarithmic contributions to the finite size scaling have not been investigated yet but will be worked out in the future. A new scan at $\kappa=0.1$ is in progress and more statistics are collected at existing data points. This will allow us to investigate the difference between the critical exponents of the Higgs-Yukawa model and the $O(4)$ model in more detail.

\smallskip

The finite temperature phase transition is found to be of second order for fermions with a physical top quark mass. The simulations for the very heavy fermions are in productions and we will be able to investigate the phase transitions there shortly. Also, the critical temperature has to be computed in both cases and must be compared to the findings of pure $O(4)$-model.

%%%%%%%%%%%%%%%%%%%%%%%%%%%%%%%%%
%
%					Acknowledgements
%
%%%%%%%%%%%%%%%%%%%%%%%%%%%%%%%%%
\section*{Acknowledgements}
This work is supported by Taiwanese
NSC via grants 100-2745-M-002-002-ASP (Academic Summit Grant),
99-2112-M-009-004-MY3, 101-2811-M-033-008, and 101-2911-I-002-509, and
by the DFG through the DFG-project Mu932/4-4, and the JSPS Grant-in-Aid for Scientific
Research (S) number 22224003. Simulations have been performed at the SGI system HLRN-II at the HLRN supercomputing service Berlin-Hannover, the PAX cluster at DESY-Zeuthen, and HPC facilities at National Chiao-Tung University and National Taiwan University. We thank the Galileo Galilei Institute for Theoretical Physics for hospitality and the INFN for the partial support during the completion of this work.

%%%%%%%%%%%%%%%%%%%%%%%%%%%%%%%%%
%
%					Bibliography
%
%%%%%%%%%%%%%%%%%%%%%%%%%%%%%%%%%
%\section{thebibliography}

\bibliographystyle{./style.bst}
\bibliography{./AHEP_refs}
%\bibliography{./poster_references}

\end{document}